\documentclass[journal,twocolumn]{IEEEtran}
\usepackage{color}
\usepackage{latexsym}
\usepackage{graphicx}
\usepackage{amsmath,amssymb}
\usepackage{verbatim,bbm}
\usepackage{amsmath}
\usepackage{amsfonts}
\usepackage{amssymb}
\providecommand{\U}[1]{\protect\rule{.1in}{.1in}}
\newtheorem{theorem}{Theorem}

\newtheorem{proposition}{Proposition}
\newtheorem{definition}{Definition}
\newtheorem{remark}{Remark}

\begin{document}

\title{Quantum data hiding in the presence of noise}

\author{Cosmo Lupo, Mark M. Wilde, 
        and Seth Lloyd
\thanks{C.~Lupo is with the
Research Laboratory of Electronics, Massachusetts Institute of Technology,
Cambridge, Massachusetts 02139, USA (email: clupo@mit.edu).}
\thanks{M.~M.~Wilde is with the 
Hearne Institute for Theoretical Physics, Department of Physics and Astronomy,
Center for Computation and Technology, Louisiana State University, Baton
Rouge, Louisiana 70803, USA (email: mwilde@lsu.edu).}
\thanks{S.~Lloyd is with the
Research Laboratory of Electronics
and the Department of Mechanical Engineering, Massachusetts Institute of Technology,
Cambridge, Massachusetts 02139, USA (email: slloyd@mit.edu).}
}


\maketitle

\begin{abstract}
When classical or quantum information is broadcast to separate receivers,
there exist codes that encrypt the encoded data such that the receivers cannot
recover it when performing local operations and classical communication, but
they can decode reliably if they bring their systems together and perform a
collective measurement. This phenomenon is known as quantum data hiding and
hitherto has been studied under the assumption that noise does not affect the
encoded systems. With the aim of applying the quantum data hiding effect in
practical scenarios, here we define the data-hiding capacity for hiding
classical information using a quantum channel. Using this notion, we establish
a regularized upper bound on the data hiding capacity of any quantum broadcast
channel, and we prove that coherent-state encodings have a strong limitation on
their data hiding rates. We then prove a lower bound on the data hiding
capacity of channels that map the maximally mixed state to the maximally
mixed state (we call these channels ``mictodiactic''---they can be seen as a generalization
of unital channels when the input and output spaces are not necessarily isomorphic) 
and argue how to extend this bound to generic channels and to more than two receivers.
\end{abstract}

\section{Introduction}

One of the primary goals of theoretical quantum information science is to
identify significant separations between the classical and quantum theories of
information. Many grand successes in this spirit have already been
achieved:\ Bell inequalities \cite{bell1964}, unconditionally secure
communication \cite{bb84}, quantum teleportation \cite{PhysRevLett.70.1895},
super-dense coding \cite{PhysRevLett.69.2881}, communication complexity
\cite{R99}, quantum data locking \cite{DHLST04}, and so on.

Another such notable example is the quantum data hiding effect
\cite{TDL01,EW02,DLT02,DHT03}. Quantum data hiding is a communication protocol
that allows for the encoding of classical or quantum information into a
two-party system such that the information can be decoded if the systems are
located in the same physical laboratory, whereas the information cannot be
decoded if the systems are located in spatially separated laboratories, even
if the laboratories are allowed to exchange classical messages. Such a task is
impossible in the classical world, simply because the exchange of a classical
message from one laboratory to the other allows for sending the entire
classical system, such that it is subsequently located in the same physical
laboratory and can thus be decoded. Arguing from the uncertainty principle of
quantum mechanics, quantum data hiding allows for the encoding of information
such that it is indistinguishable by local measurements and classical
communication and one instead requires a global measurement to distinguish the
states (i.e., information can be \textquotedblleft hidden\textquotedblright%
\ in this way).

The original proposal for quantum data hiding was an impressive observation
\cite{TDL01}, the idea being to hide information in Bell states such that it
can be recovered by a joint quantum operation but not by local operations and
classical communication. The scheme from \cite{TDL01} has a low rate of
hiding, and a quantum optical implementation was suggested. However, the
proposal rested on there not being any loss or noise in the system and was
thus impractical in several regards. Nevertheless, this striking phenomenon
prompted further research, which eventually culminated in the high-rate
schemes of \cite{HLSW04}. These schemes relied upon the use of collective
random unitary operations to hide both classical and quantum data, invoking
mathematical techniques such as the concentration of measure. The newly
proposed schemes still required no loss or noise in the encoded systems, which
exclude their application in practice. These schemes were later extended to
the case of hiding information from multiple parties \cite{HLS05}.

Even though this fascinating protocol has been discussed for some years now in
the literature and a variety of applications are now known, no one (as far as
we are aware) has yet to consider the most pertinent practical question: Is
\textit{quantum data hiding possible in the presence of noise and loss, and if
so, what is its performance?} Here, we address this fundamental question using
the language of quantum Shannon theory and recently developed techniques that
have established quantum data locking protocols in the presence of noise
\cite{GHKLLSTW13,LWL14,LL14,LL15,LL15a} (see also \cite{W14} for another
advance in the same direction). The channels of primary interest to us are
those which have the property of mapping the maximally mixed state in the input
space into the maximally mixed state in the output space. We called these channels
{\it mictodiactic} (literally, Greek for ``mixtures passing through'') as they preserve the maximally mixed state. Mictodiactic channels
coincide with unital channels if the input and output space have the same
dimension.
We focus on these channels mainly because they are relatively simple to analyze.

The main development of this paper begins with a formal definition of the data
hiding capacity of a quantum broadcast channel. From there, we establish a
regularized upper bound on the data hiding capacity of any broadcast channel,
and we then show that coherent-state protocols cannot offer a high rate of
data hiding (this is perhaps to be expected, given that coherent states are
\textquotedblleft pseudo-classical\textquotedblright\ and in light of our
prior work in which we showed that they do not offer high rates for quantum
data locking \cite{GHKLLSTW13}). We next establish a lower bound on the data
hiding capacity of a mictodiactic quantum broadcast channel and argue that this
scheme can be extended to generic channels and to a multipartite scenario. 
The scheme involves random coding arguments, and we invoke concentration of measure bounds in order to
establish security of the scheme.

\section{Preliminaries}

Here we recall some basic facts before beginning the main development. A
quantum state is represented by a density operator, which is a positive
semi-definite operator acting on a Hilbert space $\mathcal{H}$\ and with trace
equal to one. A multipartite quantum state acts on a tensor product of Hilbert
spaces. A classical-quantum state $\rho_{XB}$\ has the following form:%
\begin{equation}
\rho_{XB}\equiv\sum_{x}p_{X}(  x)  \vert x\rangle
\langle x\vert _{X}\otimes\rho_{B}^{x},
\end{equation}
where $p_{X}$ is a probability distribution, $\left\{  \left\vert
x\right\rangle _{X}\right\}  $ is an orthonormal basis, and $\left\{  \rho
_{B}^{x}\right\}  $ is a set of quantum states. A quantum channel is modeled
as a completely positive, trace-preserving linear map, taking operators acting
on one Hilbert space to operators acting on another one. 
A quantum channel is
unital if it preserves the identity operator (note that the input and output
space of such a channel must have the same dimension). 
A quantum measurement
is a special kind of quantum channel which accepts a quantum input system and
outputs a classical system. That is, it can be described as follows:%
\begin{equation}
\mathcal{M}(  \rho)  =\sum_{y}\operatorname{Tr}\left\{  \Lambda
^{y}\rho\right\}  \vert y\rangle \langle y\vert ,
\end{equation}
where $\Lambda^{y}\geq0$ for all $y$, $\sum_{y}\Lambda^{y}=I$, and $\left\{
\vert y\rangle \right\}  $ is some known orthonormal basis. The
collection $\left\{  \Lambda^{y}\right\}  $ is called a positive
operator-valued measure (POVM). A quantum instrument is a quantum channel that
accepts a quantum input and outputs a quantum system and a classical system.

A quantum broadcast channel is defined as a quantum channel that accepts one
input quantum system $A$ and has two output quantum systems $B$ and $C$
\cite{YHD2006}. Formally, it is a completely positive trace preserving linear
map from the space of operators that act on the Hilbert space for system $A$
to the space of operators acting on the tensor-product Hilbert space for
systems $B$ and $C$. An important example of such a channel is one that mixes
an input mode with two vacuum states at two beamsplitters such that the two
receivers obtain two of the beamsplitter outputs while the environment obtains
the other mode \cite{GSE07,STW15}. One could incorporate noise effects in
addition to loss by mixing with thermal states instead of vacuum states.


The class of local operations and classical communication (LOCC)\ consists of
compositions of the following operations:

\begin{enumerate}
\item Alice performs a quantum instrument, which has both a quantum and
classical output. She forwards the classical output to Bob, who then performs
a quantum channel conditioned on the classical data received.

\item The situation is reversed, with Bob performing the initial instrument,
who forwards the classical data to Alice, who then performs a quantum
channel conditioned on the classical data.
\end{enumerate}
An LOCC\ measurement is a measurement channel that can be implemented by means
of local operations and classical communication. Every LOCC\ channel can be
written in the following form (as a separability preserving channel):%
\begin{equation}
\mathcal{L}_{AB}(  \rho_{AB})  =\sum_{x}\left(  C_{A}^{x}\otimes
D_{B}^{x}\right)  \rho_{AB}\left(  C_{A}^{x}\otimes D_{B}^{x}\right)  ^{\dag},
\end{equation}
such that $\sum_{x}\left(  C_{A}^{x}\otimes D_{B}^{x}\right)  ^{\dag}\left(
C_{A}^{x}\otimes D_{B}^{x}\right)  =I_{AB}$. However, not every channel of the
form above is an LOCC\ channel \cite{PhysRevA.59.1070}.

The trace norm of an operator $A$ is defined as $\left\Vert A\right\Vert
_{1}\equiv\operatorname{Tr}[\sqrt{A^{\dag}A}]$, and it induces the trace
distance $\left\Vert \rho-\sigma\right\Vert _{1}$ as a measure of
distinguishability between two quantum states $\rho$ and $\sigma$. The von
Neumann entropy of a state $\rho$ on system $A$ is equal to $S(
A)  _{\rho}\equiv S(  \rho)  \equiv-\operatorname{Tr}%
[\rho\log\rho]$. The quantum mutual information of a bipartite state
$\rho_{AB}$ is defined as%
\begin{equation}
I(  A;B)  _{\rho}\equiv S(  A)  _{\rho}+S(
B)  _{\rho}-S(  AB)  _{\rho}.
\end{equation}
The LOCC accessible information of a classical-quantum-quantum state
$\rho_{MBC}$ is as follows:%
\begin{equation}
I_{\operatorname{acc},\operatorname{LOCC}}(  M;BC)  _{\rho}%
\equiv\max_{\mathcal{L}_{BC\rightarrow\hat{M}}\in\operatorname{LOCC}}%
I(M;\hat{M})_{\omega},
\end{equation}
where the mutual information on the RHS\ is with respect to the following
classical-classical state:%
\begin{equation}
\omega_{M\hat{M}}\equiv\mathcal{L}_{BC\rightarrow\hat{M}}(  \rho
_{MBC})  ,
\end{equation}
and $\mathcal{L}_{BC\rightarrow\hat{M}}$ is an LOCC\ measurement channel.

\section{Quantum data hiding capacity}

In a quantum data-hiding protocol, the sender Alice communicates classical or
quantum information to two spatially separated receivers Bob and Charlie via a
quantum broadcast channel $\mathcal{N}_{A\rightarrow BC}$. The protocol
satisfies the ``correctness property'' if Bob and Charlie can decode reliably
when allowed to apply a joint quantum measurement (i.e., if they are located
in the same laboratory). On the other hand, the ``data-hiding/security
property'' is that the transmitted information cannot be accessed by Bob and
Charlie when they are restricted to performing local operations and classical
communication (i.e., if they are in different laboratories connected by a
classical communication channel). When Alice sends classical information, we
call this a \textit{bit-hiding protocol} and define it as follows:

\begin{definition}
[Bit-hiding protocol]An $(n,M,\delta,\varepsilon)$ bit-hiding protocol for a
quantum broadcast channel $\mathcal{N}_{A\rightarrow BC}$ consists of 
a collection of input states $\{ \rho(x) \}_{x=1,\dots,M}$, and a decoding
measurement satisfying the following properties:

\begin{itemize}
\item \textit{(Correctness)} It is possible to decode with high average success probability, that is, 
\begin{equation}
\frac{1}{M}\sum_{x}\operatorname{Tr}\left[  \Lambda_{B^{n}C^{n}}%
^{x}\mathcal{N}_{A\rightarrow BC}^{\otimes n}(\rho(x))\right]  \geq 1-\delta\,,
\label{cor}%
\end{equation}
where $\{\Lambda_{B^{n}C^{n}}^{x}\}$ is the POVM  associated to the
decoding measurement.

\item \textit{(Security)} For all LOCC measurements $\mathcal{M}%
_{\operatorname{LOCC}}$ on the bipartite system $B^{n}C^{n}$, there exists a
state $\sigma_{B^{n}C^{n}}$ such that%
\begin{equation}
\frac{1}{M}\sum_{x}\left\Vert \mathcal{M}_{\operatorname{LOCC}}(
\mathcal{N}_{A\rightarrow BC}^{\otimes n}(\rho(x))-\sigma_{B^{n}C^{n}})
\right\Vert _{1}\leq\varepsilon\,. \label{sec}%
\end{equation}

\end{itemize}
\end{definition}

\begin{remark}
Due to the convexity of the trace norm, it is sufficient to prove the security
property against rank-one LOCC measurements.
\end{remark}

We can now define the bit-hiding capacity of a quantum broadcast channel
$\mathcal{N}_{A\rightarrow BC}$:

\begin{definition}
[Bit-hiding capacity]A bit-hiding rate $R$ is achievable for a quantum
broadcast channel $\mathcal{N}_{A\rightarrow BC}$ if for all $\delta
,\varepsilon, \zeta\in\left(  0,1\right)  $ and sufficiently large $n$, there
exists an $(n,M,\delta,\varepsilon)$ bit-hiding protocol such that $\frac
{1}{n}\log{M}\geq R-\zeta$. The bit-hiding capacity $\kappa(\mathcal{N})$ of
$\mathcal{N}$ is equal to the supremum of all achievable bit-hiding rates.
\end{definition}

In this paper we focus on bit-hiding protocols. One can also define
qubit-hiding protocols, aimed at hiding quantum information (see
\cite{HLSW04}). Similarly, one could define a notion of qubit-hiding capacity
of a quantum channel. Additionally, these definitions immediately generalize
to the case of a channel with one sender and an arbitrary number of receivers.

\section{Regularized upper bound on quantum data hiding capacity}

We begin by establishing a regularized upper bound on the bit-hiding capacity
of any channel:

\begin{theorem}
The bit-hiding capacity $\kappa(\mathcal{N})$\ of a quantum channel
$\mathcal{N}$\ is bounded from above as follows:%
\begin{equation}
\kappa(\mathcal{N})\leq\lim_{n\rightarrow\infty}\frac{1}{n}\kappa^{\left(
u\right)  }(\mathcal{N}^{\otimes n}),
\end{equation}
where%
\begin{align}
\kappa^{(  u)  }(\mathcal{N})  &  \equiv\max_{\left\{  p_{X}(
x)  ,\rho_{x}\right\}  }\left[  I(  X;BC)  _{\rho
}-I_{\operatorname{acc},\operatorname{LOCC}}(  X;BC)  _{\rho
}\right]  ,\nonumber\\
\rho_{XBC}  &  \equiv\sum_{x}p_{X}(  x)  \vert x\rangle
\langle x\vert _{X}\otimes\mathcal{N}_{A\rightarrow BC}(
\rho_{x})  .
\end{align}

\end{theorem}

\begin{IEEEproof}
This upper bound follows by using the definition of a bit-hiding protocol and
a few well known facts. Let us consider an $\left(  n,M,\delta,\varepsilon
\right)  $ bit hiding protocol.
By a simple reduction (see Appendix~\ref{sec:trace-dist-err-prob}), the correctness criterion in (\ref{cor})
translates to the following criterion:
\begin{equation}
\frac{1}{2}\left\Vert \overline{\Phi}_{MM^{\prime}}-\omega_{MM^{\prime}%
}\right\Vert _{1}\leq\delta, \label{eq:trace-dist-err-prob-init}
\end{equation}
where $\overline{\Phi}_{MM^{\prime}}$ is a maximally correlated state, defined
as%
\begin{equation}
\overline{\Phi}_{MM^{\prime}}\equiv\frac{1}{M}\sum_{x}\vert
x\rangle \langle x\vert _{M}\otimes\vert x\rangle
\langle x\vert _{M^{\prime}},
\end{equation}
and $\omega_{MM^{\prime}}$ is defined from the probability distribution on the
LHS\ of (\ref{cor}) as%
\begin{equation}
\sum_{x,x^{\prime}}\frac{1}{M}\operatorname{Tr}\left[  \Lambda_{B^{n}C^{n}%
}^{x^{\prime}}\mathcal{N}_{A\rightarrow BC}^{\otimes n}(\rho(x))\right]
\vert x\rangle \langle x\vert _{M}\otimes\vert
x^{\prime}\rangle \langle x^{\prime}\vert _{M^{\prime}}.
\end{equation}
Then%
\begin{align}
\log M  &  =I(  M;M^{\prime})  _{\overline{\Phi}}\\
&  \leq I(  M;M^{\prime})  _{\omega}+f(  n,\delta) \\
&  \leq I(  M;B^{n}C^{n})  _{\omega}+f(  n,\delta)  .
\end{align}
The first inequality is an application of the Alicki-Fannes continuity of
entropy inequality \cite{AF04}, with $f(  n,\delta)  $ a function
such that $\lim_{\delta\rightarrow0}\lim_{n\rightarrow\infty}\frac{1}%
{n}f(  n,\delta)  =0$. The second inequality is a consequence of
the Holevo bound \cite{Ho73}, where we have denoted the state after the
channel (but before the decoding measurement)
again by $\omega$. As a consequence of the security criterion in
\eqref{sec} and the Alicki-Fannes inequality, we can conclude that%
\begin{equation}
I_{\operatorname{acc},\operatorname{LOCC}}(  M;B^{n}C^{n})  \leq
g(  n,\varepsilon)  ,
\end{equation}
with $g(  n,\varepsilon)  $ a function such that
$
\lim_{\varepsilon\rightarrow0}\lim_{n\rightarrow\infty}\frac{1}{n}g(
n,\varepsilon)  =0.
$
So this gives%
\begin{multline}
\log M\leq I(  M;B^{n}C^{n})  -I_{\operatorname{acc}%
,\operatorname{LOCC}}(  M;B^{n}C^{n}) \\
+f(  n,\delta)  +g(  n,\varepsilon)  .
\label{eq:gen-upp-bound}%
\end{multline}
We finally optimize over all possible channel inputs, divide by $n$, take the
limit as $n\rightarrow\infty$ and then as $\varepsilon,\delta\rightarrow0$ to
establish the statement of the theorem.
\end{IEEEproof}

\section{Upper bound for coherent-state data-hiding protocols}

Here we show that data-hiding schemes making use of coherent-state encodings
are highly limited in terms of the rate at which they can hide information. We
assume that the mean input photon number is less than $N_{S} \in(0,\infty)$
and Alice is connected to Bob and Charlie by a bosonic broadcast channel
\cite{GSE07}, in which Alice has access to one input port of a beamsplitter,
the vacuum is injected into the other input port, and Bob and Charlie have
access to the outputs of the beamsplitter. The main idea behind the bound that
we prove here is that the classical capacity of the pure-loss bosonic channel
is limited from above by $g(N_{S})$ \cite{PhysRevLett.92.027902}, where $g(  x)
\equiv\left(  x+1\right)  \log_{2}\left(  x+1\right)  -x\log_{2}x$. At the
same time, if Bob and Charlie perform heterodyne detection on their outputs
and coordinate their results, the rate at which they can decode information is
equal to $\log_{2}(1+N_{S})$. Our proof of the following theorem makes this
intuition rigorous. Notably, this bound is the same as that  found in
\cite{GHKLLSTW13} for the strong locking capacity when using coherent-state encodings.

\begin{theorem}
The quantum data-hiding capacity of a bosonic broadcast channel when
restricting to coherent-state encodings with mean photon number $N_{S}$ is
bounded from above by $g(N_{S})-\log_{2}(1+N_{S}) \leq\log_{2}(e) \approx1.45$.
\end{theorem}

\begin{IEEEproof}
Consider a quantum data-hiding scheme consisting of coherent-state
codewords:\ $\left\{  \vert \alpha^{n}(  x,k)  \rangle
\right\}  _{x,k}$, where $\left\vert \alpha^{n}\right\rangle \equiv\left\vert
\alpha_{1}\right\rangle \otimes\left\vert \alpha_{2}\right\rangle
\otimes\cdots\otimes\left\vert \alpha_{n}\right\rangle $ and such that the
mean photon number of the scheme is less than $N_{S}\in(0,\infty)$. Then the
quantum codeword transmitted for message $x$\ is as follows:%
\begin{equation}
\rho(  x)  \equiv\frac{1}{K}\sum_{k}\vert \alpha^{n}(
x,k)  \rangle \langle \alpha^{n}(  x,k)
\vert .
\end{equation}
So the classical-quantum state corresponding to the transmitted state is as
follows:%
\begin{multline}
\omega_{MKB^{n}C^{n}}\equiv  \\
\frac{1}{MK}\sum_{x,k}\vert x\rangle
\langle x\vert _{M}\otimes\vert k\rangle \langle
k\vert _{K}  
 \otimes\left\vert \sqrt{\eta}\alpha^{n}(  x,k)  \right\rangle
\left\langle \sqrt{\eta}\alpha^{n}(  x,k)  \right\vert _{B^{n}}  \\
 \otimes\left\vert \sqrt{1-\eta}\alpha^{n}(  x,k)  \right\rangle
\left\langle \sqrt{1-\eta}\alpha^{n}(  x,k)  \right\vert _{C^{n}},
\end{multline}
where $\left\vert \sqrt{\eta}\alpha^{n}\right\rangle $ is a shorthand for
$\left\vert \sqrt{\eta}\alpha_{1}\right\rangle \otimes\left\vert \sqrt{\eta
}\alpha_{2}\right\rangle \otimes\cdots\otimes\left\vert \sqrt{\eta}\alpha
_{n}\right\rangle $.

In order to obtain our upper bound, we suppose that Bob performs heterodyne
detection, forwards the results to Charlie, who also performs heterodyne
detection and coordinates the results to decode. Now begin from the upper
bound in \eqref{eq:gen-upp-bound}:%
\begin{align}
& I(  M;B^{n}C^{n})  _{\omega}-I_{\text{acc,LOCC}}(
M;B^{n}C^{n})  _{\omega} \nonumber \\
&  \leq I(  M;B^{n}C^{n})  _{\omega}-I_{\text{het,LO}}(
M;B^{n}C^{n})  _{\omega}\\
&  = I(  MK;B^{n}C^{n})  _{\omega}-I(  K;B^{n}C^{n}|M)
_{\omega} \nonumber \\
& \quad - \left[  I_{\text{het,LO}}(  MK;B^{n}C^{n})  _{\omega
}-I_{\text{het,LO}}(  K;B^{n}C^{n}|M)  _{\omega}\right]  \\
&  =I(  MK;B^{n}C^{n})  _{\omega}-I_{\text{het,LO}}(
MK;B^{n}C^{n})  _{\omega} \nonumber \\
& \quad -\left[  I(  K;B^{n}C^{n}|M)  _{\omega}-I_{\text{het,LO}%
}(  K;B^{n}C^{n}|M)  _{\omega}\right]  \\
&  \leq I(  MK;B^{n}C^{n})  _{\omega}-I_{\text{het,LO}}(
MK;B^{n}C^{n})  _{\omega}\\
&  \leq\max_{p_{Y}(  y)  }\left[  I(  Y;B^{n}C^{n})
_{\tau}-I_{\text{het,LO}}(  Y;B^{n}C^{n})  _{\tau}\right]  \\
&  \leq n\max_{p_{Y}(  y)  }\left[  I(  Y;BC)  _{\sigma
}-I_{\text{het,LO}}(  Y;BC)  _{\sigma}\right]
.\label{eq:coh-state-last-line}%
\end{align}
The first inequality follows by picking the LOCC measurement to be the
classically coordinated heterodyne detection mentioned above. The first
equality follows from the chain rule for conditional mutual information, and
the second equality is a rewriting. The second inequality follows because
$I(  K;B^{n}C^{n}|M)  _{\omega}-I_{\text{het,LO}}\left(
K;B^{n}C^{n}|M\right)_{\omega}\geq0$, which is a consequence of data
processing:\ the classical heterodyne detection measurement outcomes are from
measuring systems $B^{n}C^{n}$. The second-to-last inequality follows by
optimizing over all input distributions and the information quantities are
evaluated with respect to a state of the following form:%
\begin{multline}
\tau_{YB^{n}C^{n}}\equiv\sum_{y}p_{Y}(  y)  \vert
y\rangle \langle y\vert _{Y}\otimes\left\vert \sqrt{\eta
}\alpha^{n}(  y)  \right\rangle \left\langle \sqrt{\eta}\alpha
^{n}(  y)  \right\vert _{B^{n}} \\
\otimes\left\vert \sqrt{1-\eta}\alpha^{n}(  y)  \right\rangle
\left\langle \sqrt{1-\eta}\alpha^{n}(  y)  \right\vert _{C^{n}}.
\end{multline}
The last inequality follows by realizing that the difference between the
mutual informations can be understood as being equal to the private
information of a quantum wiretap channel in which the state is prepared for
the receiver while the heterodyned version of this state (a classical
variable) is prepared for the eavesdropper. Such a quantum wiretap channel has
pure product input states (they are coherent states) and it is degraded. Thus,
we can apply Theorem 35 from \cite{GHKLLSTW13}\ to conclude that this private
information is subadditive. The information quantities in the last line are
with respect to a state of the following form:%
\begin{multline}
\sigma_{YBC}\equiv\sum_{y}p_{Y}(  y)  \vert y\rangle
\langle y\vert _{Y}\otimes\left\vert \sqrt{\eta}\alpha\left(
y\right)  \right\rangle \left\langle \sqrt{\eta}\alpha(  y)
\right\vert _{B} \\ 
\otimes\left\vert \sqrt{1-\eta}\alpha(  y)  \right\rangle
\left\langle \sqrt{1-\eta}\alpha(  y)  \right\vert _{C}.
\end{multline}

Now by a development nearly identical to that given in Eqs.~(34)-(42) of
\cite{GHKLLSTW13}, we can conclude that a circularly symmetric, Gaussian
mixture of coherent states with variance $N_{S}$ optimizes the quantity in
(\ref{eq:coh-state-last-line}). For such a distribution, the quantity
$I(  Y;BC)  _{\sigma}$ evaluates to%
\begin{equation}
I(  Y;BC)  _{\sigma}=g(  N_{S})  .
\end{equation}
We now evaluate the quantity $I_{\text{het,LO}}(  Y;BC)  _{\sigma}%
$. Consider that the output for Bob is a random variable $\sqrt{\eta}%
\alpha+z_{B}$, where $\alpha$ is a zero-mean Gaussian random variable
(RV)\ with variance $N_{S}$ and $z_{B}$ is a zero-mean Gaussian RV\ with
variance $1$. The output for Charlie is a RV $\sqrt{1-\eta}\alpha+z_{C}$,
where $\alpha$ is the same Gaussian RV\ as above and $z_{C}$ is a zero-mean
Gaussian RV\ with variance $1$ (note that $\alpha$, $z_{B}$, and $z_{C}$ are
independent RVs). The covariance matrix for the real part of these RVs is as
follows:%
\begin{equation}%
\begin{bmatrix}
\eta N_{S}/2+1/2 & \sqrt{\eta\left(  1-\eta\right)  }N_{S}/2\\
\sqrt{\eta\left(  1-\eta\right)  }N_{S}/2 & \left(  1-\eta\right)  N_{S}/2+1/2
\end{bmatrix}
,
\end{equation}
with determinant equal to%
\begin{equation}
\frac{1}{4}(  N_{S}+1)  .
\end{equation}
The determinant of the covariance matrix of the real parts of the RVs $z_{B}$
and $z_{C}$ is equal to $1/4$. By modeling the real and imaginary components
as two independent parallel channels and plugging into the Shannon formula for
the capacity of a Gaussian channel, we find that%
\begin{equation}
I_{\text{het,LO}}(  Y;BC)  _{\sigma}=\log(  N_{S}+1)  .
\end{equation}
So we can finally conclude the upper bound on the data hiding capacity of
coherent-state schemes.
\end{IEEEproof}

\section{Lower bound on quantum data hiding capacity}

Here we prove a lower bound for the bit-hiding capacity of mictodiactic quantum
broadcast channels. We consider a channel $\mathcal{N}%
_{A\rightarrow BC}$ from an input quantum system of dimensions $\dim{A}=d_{A}$
to the output systems with dimensions $\dim{B}=d_{B}$ and $\dim{C}=d_{C}$. 
A mictodiactic channel is defined from the following property:
\begin{equation}
\mathcal{N}_{A\rightarrow BC}(I_{A}/d_{A}%
)=I_{BC}/(d_{B}d_{C}),
\end{equation}
where $I_{A}$ and $I_{BC}$ denote the identity
operators acting on the input and output spaces, respectively. 

\begin{theorem}
\label{main} The bit-hiding capacity $\kappa(  \mathcal{N})  $ of a
 mictodiactic broadcast channel $\mathcal{N}_{A\rightarrow BC}$, with
$\dim{A}=d_{A}$, $\dim{B}=d_{B}$ and $\dim{C}=d_{C}$
is bounded from below as follows:%
\begin{equation}
\kappa(  \mathcal{N})  \geq\kappa^{\left(  l\right)  }%
(\mathcal{N})\equiv\chi(\mathcal{N})-\log{d_{+}}-\log{\gamma}\,.
\end{equation}
where
\begin{equation}
\chi(\mathcal{N})=S\left(  \int d\psi\ \mathcal{N}(\psi)\right)  -\int
d\psi\ S(\mathcal{N}(\psi))\,
\end{equation}
is the Holevo information of the channel computed for a uniform ensemble of
input states ($d\psi$ denotes the uniform measure on the sphere of unit
vectors), and $d_{+}=\max{\{d_{B},d_{C}\}}$. The parameter $\gamma$ is given
by
\begin{equation}
\gamma = \frac{2  d_{B}^2 d_{C}^2 }{d_A(d_{A} + 1)} \, 
\left\Vert (\mathcal{N}\otimes\mathcal{N} )(P_\mathrm{sym})\right\Vert _{\infty}
\,, \label{gammadef}
\end{equation}
where $\Vert\,\cdot\,\Vert_{\infty}$ is the $\infty$-norm and $P_\mathrm{sym}$ is the
projector onto the symmetric subspace.

\end{theorem}

A proof is given in the next section.

As an example consider the $d$-dimensional depolarizing channel, with
$d=d_{A}=d_{B}d_{C}$:
\begin{equation}
\mathcal{N}(X) = p X + (1-p)\operatorname{Tr}(X) I/d.
\end{equation}
Let us first compute the Holevo information for the uniform distribution of
input states. A straightforward calculation yields
\begin{multline}
\chi(\mathcal{N}) = \log{d} 
+\left(  p+\frac{1-p}{d}\right)  \log{\left(  p+\frac{1-p}{d}\right)  } \\
+ \left( d-1 \right) \left(  \frac{1-p}{d}\right)  \log{\left( \frac{1-p}{d}\right)  } \, .
\end{multline}
Then consider the density matrix 
$\frac{P_\mathrm{sym}}{\mathrm{Tr}(P_\mathrm{sym})} = 2\frac{P_\mathrm{sym}}{d(d+1)}$, which is
mapped by $(\mathcal{N}\otimes\mathcal{N})$ into 
\begin{equation}
(\mathcal{N}\otimes\mathcal{N})\left( \frac{P_\mathrm{sym}}{\mathrm{Tr}(P_\mathrm{sym})} \right) 
= p^{2}\frac{P_\mathrm{sym}}{\mathrm{Tr}(P_\mathrm{sym})} 
+ (1-p^{2})\frac{I\otimes I}{d^{2}}\,,
\end{equation}
where we have used the fact that
the partial trace of $\frac{P_\mathrm{sym}}{\mathrm{Tr}(P_\mathrm{sym})}$ is the
maximally mixed state $I/d$.
We hence have
\begin{align}
& \!\!\!\!\left\Vert (\mathcal{N}\otimes\mathcal{N} )(P_\mathrm{sym})\right\Vert _{\infty} \nonumber \\
& = \mathrm{Tr}(P_\mathrm{sym}) \left\Vert (\mathcal{N}\otimes\mathcal{N})\left( \frac{P_\mathrm{sym}}{\mathrm{Tr}(P_\mathrm{sym})} \right) \right\Vert _{\infty} \\
& = \frac{d(d+1)}{2} \left( \frac{2p^2}{d(d+1)} + \frac{1-p^2}{d^2} \right) \\
& = \frac{d+1}{2d} + p^2 \left( 1 - \frac{d+1}{2d} \right)\, .
\end{align}
Finally, from (\ref{gammadef}), putting $d_{B}d_{C} = d$ we obtain
\begin{equation}
\gamma = 1 + p^{2} \left( \frac{2d}{d+1} - 1\right) \, .
\end{equation}

For the case of unital channels, which are contractive with respect to the
Schatten norms \cite{PWPR06}, we have $\left\Vert (\mathcal{N}\otimes
\mathcal{N})(P_\mathrm{sym})\right\Vert _{\infty}\leq\left\Vert P_\mathrm{sym}\right\Vert
_{\infty}=1$, from which we obtain the following looser bound:
\begin{equation}
\kappa^{\left(  l\right)  }(\mathcal{N})\geq\chi(\mathcal{N})-\log{d_{+}}%
-\log{\frac{2d}{d+1}}\,.
\end{equation}



\subsection{Proof of Theorem \ref{main}}

We now provide a proof for Theorem \ref{main}. We generate a bit-hiding code for $n$ uses of the quantum broadcast channel
$\mathcal{N}_{A\rightarrow BC}$, by employing the well known method of random
coding. For $x=1,\dots,M$, pick a collection of $M$ pure quantum states at
random, each having the following form:%
\begin{equation}
|\psi(x)\rangle=\bigotimes\limits_{j=1}^{n}|\psi_{j}(x)\rangle.
\end{equation}
These random unit vectors are each generated by sampling $|\psi_{j}(x)\rangle$
independently from the uniform distribution over the unit sphere in
$\mathbb{C}^{d}$. A simple way of doing so is to single out a fiducial unit
vector $\left\vert 0\right\rangle \in\mathbb{C}^{d}$ and pick a unitary
$V_{xj}$ at random according to the Haar measure on $SU(d)$, in order to
generate $|\psi_{j}(x)\rangle=V_{xj}\left\vert 0\right\rangle $. Let%
\begin{equation}
V_{x}=\bigotimes\limits_{j=1}^{n}V_{xj}.
\end{equation}
We then pick $Kn$ qudit unitaries $U_{kj}$ i.i.d.\ from the Haar measure on
$SU(d)$. The inputs to the channel have the following form:%
\begin{equation}
\rho(x)=\frac{1}{K}\sum_{k=1}^{K} U_{k} \, \psi(x) \, U_{k}^{\dag}\,, \label{rando}%
\end{equation}
where
\begin{equation}
U_{k}=\bigotimes\limits_{j=1}^{n}U_{kj}.
\end{equation}

We introduce the notation $\mathbf{V}=(V_{1},\dots,V_{M})$ to denote the
$M$-tuple of $n$-local qudit unitaries and the notation $\mathbf{U}%
=(U_{1},\dots,U_{K})$ to denote the $K$-tuple of $n$-local qudit unitaries,
and we further define $\mathcal{R}_{\mathbf{U}}$ as follows:%
\begin{equation}
\mathcal{R}_{\mathbf{U}}(\sigma)=\frac{1}{K}\sum_{k=1}^{K}U_{k}\,\sigma
\,U_{k}^{\dag}\,.
\end{equation}
We also define the random variables $\boldsymbol{\mathcal{V}}$ and
$\boldsymbol{\mathcal{U}}$, taking values on the set of $M$-tuples
$\mathbf{V}=(V_{1},\dots,V_{M})$ and $K$-tuples $\mathbf{U}=(U_{1},\dots
,U_{K})$, respectively, where each qudit unitary is statistically independent
and uniformly distributed according to the distribution induced by the Haar
measure on $SU(d)$.

Given the input codewords in (\ref{rando}), the state at the output of $n$
channel uses is as follows:%
\begin{align}
\mathcal{N}^{\otimes n}(\rho(x))  &  =\mathcal{N}^{\otimes n}(\mathcal{R}%
_{\mathbf{U}}( \psi(x) ))\\
&  = \mathcal{N}^{\otimes n} ( \mathcal{R}_{\mathbf{U}}(V_{x}|0\rangle\langle0
| V_{x} ) ) \, .
\end{align}

We split the proof into two parts: first we prove the correctness property and
then the security property.

\subsubsection{Correctness}

The proof of the correctness property has a rather standard form, but we
provide it for completeness. Let $\Pi_{\mathcal{N}(  \pi)
}^{n,\delta_{0}}$ denote the $\delta_{0}$-weakly typical projection for the
tensor-power state $\left[  \mathcal{N}(  \pi)  \right]  ^{\otimes
n}$, where $\pi$ denotes the maximally mixed state and $\delta_{0}>0$ (see,
e.g., \cite{W13} for a definition and properties). This projector has the
following properties which hold for all $\varepsilon_{0}\in\left(  0,1\right)
$ and sufficiently large $n$:%
\begin{align}
\operatorname{Tr}(\Pi_{\mathcal{N}(  \pi)  }^{n,\delta_{0}}\left[
\mathcal{N}(  \pi)  \right]  ^{\otimes n}) &  \geq1-\varepsilon
_{0},\label{eq:uncond-typ}\\
\Pi_{\mathcal{N}(  \pi)  }^{n,\delta_{0}}\left[  \mathcal{N}(
\pi)  \right]  ^{\otimes n}\Pi_{\mathcal{N}(  \pi)
}^{n,\delta_{0}} &  \leq2^{-\left[  S\left(  \mathcal{N}(  \pi)
\right)  -\delta_{0}\right]  }\Pi_{\mathcal{N}(  \pi)  }%
^{n,\delta_{0}}.
\end{align}

Let a spectral decomposition of the state $\mathcal{N}^{\otimes n}(U_{k}%
\,\psi(x)\,U_{k}^{\dag})$ be as follows:%
\begin{equation}
\mathcal{N}^{\otimes n}(U_{k}\,\psi(x)\,U_{k}^{\dag})=\sum_{y^{n}}p\left(
y^{n}|k,x\right)  \left\vert y_{k,x}^{n}\right\rangle \left\langle y_{k,x}%
^{n}\right\vert ,
\end{equation}
where the eigenvalues $p\left(  y^{n}|k,x\right)  $ form a product
distribution because the state is tensor product. Let $\Pi_{k,x}^{n,\delta
_{1}}$ denote the $\delta_{1}$-weak conditionally typical projection
corresponding to the state $\mathcal{N}^{\otimes n}(U_{k}\,\psi(x)\,U_{k}%
^{\dag})$, defined as the projection onto the following subspace:%
\begin{equation}
\text{span}\left\{  \left\vert y_{k,x}^{n}\right\rangle :\left\vert -\frac
{1}{n}\log p\left(  y^{n}|k,x\right)  -S_{\ast}\right\vert \leq\delta
_{1}\right\}  .
\end{equation}
where $\delta_{1}>0$ and $S_{\ast}=\int d\psi\ S(  \mathcal{N}(
\psi)  )  $ (see, e.g., \cite{W13} for more details). These
projectors have the following property, which holds for all $\varepsilon_{1}%
\in\left(  0,1\right)  $ and sufficiently large $n$:%
\begin{equation}
\mathbb{E}_{U_{k},V_{x}}\left\{  \operatorname{Tr}\left(  \Pi_{k,x}%
^{n,\delta_{1}}\mathcal{N}^{\otimes n}(U_{k}\psi(x)U_{k}^{\dag})\right)
\right\}  \geq1-\varepsilon_{1}\label{eq:typicality}.
\end{equation}
(In fact, both $\varepsilon_{0}$ and $\varepsilon_{1}$ can be taken to be
exponentially decreasing in$~n$, following from an application of the Chernoff
bound.) So this means that we can build a reliable decoding measurement from
the set $\left\{  \Pi_{k,x}\right\}  $\ of projectors. Another property of the
conditionally typical projector is the following bound on the size of each
projector:%
\begin{equation}
\operatorname{Tr}(\Pi_{k,x}^{n,\delta_{1}})\leq2^{n\left[  S_{\ast}+\delta
_{1}\right]  }.
\end{equation}

We then consider a \textquotedblleft square-root\textquotedblright\ decoding
measurement defined by the following POVM elements:%
\begin{align}
\Lambda_{k,x} &  \equiv\left(  \sum_{k^{\prime},x^{\prime}}\Gamma_{k^{\prime
},x^{\prime}}\right)  ^{-1/2}\Gamma_{k,x}\left(  \sum_{k^{\prime},x^{\prime}%
}\Gamma_{k^{\prime},x^{\prime}}\right)  ^{-1/2}\,,\\
\Gamma_{k,x} &  \equiv\Pi_{\mathcal{N}(  \pi)  }^{n,\delta_{0}}%
\Pi_{k,x}^{n,\delta_{1}}\Pi_{\mathcal{N}(  \pi)  }^{n,\delta_{0}}.
\end{align}
The average probability of an error when decoding both $x$ and $k$ is as
follows:%
\begin{equation}
\bar{p}_{\operatorname{err}}=1-\frac{1}{MK}\sum_{k,x}\operatorname{Tr}%
(\Lambda_{k,x}\mathcal{N}^{\otimes n}(\psi_{k}(x))),\label{eq:avg-err-prob}%
\end{equation}
where we have introduced the shorthand $\psi_{k}(x)\equiv U_{k}\psi
(x)U_{k}^{\dag}$. Recall the Hayashi-Nagoaka operator inequality \cite{HN03}%
\begin{equation}
I-\left(  P+Q\right)  ^{-1/2}P\left(  P+Q\right)  ^{-1/2}\leq2\left(
I-P\right)  +4Q,
\end{equation}
which holds for $0\leq P\leq I$ and $Q\geq0$. Picking
\begin{equation}
P=\Gamma_{k,x},\ \ \ \ \ Q=\sum_{\left(  k^{\prime},x^{\prime}\right)
\neq\left(  k,x\right)  }\Gamma_{k^{\prime},x^{\prime}},
\end{equation}
we can apply this to (\ref{eq:avg-err-prob}) to bound it from above as%
\begin{multline}
\bar{p}_{\operatorname{err}}\leq\frac{2}{MK}\sum_{k,x}\operatorname{Tr}%
(\left[  I-\Gamma_{k,x}\right]  \mathcal{N}^{\otimes n}(\psi_{k}(x))) \\
+\frac{4}{MK}\sum_{k,x}\sum_{\left(  k^{\prime},x^{\prime}\right)  \neq\left(
k,x\right)  }\operatorname{Tr}(\Gamma_{k^{\prime},x^{\prime}}\mathcal{N}%
^{\otimes n}(\psi_{k}(x))).
\end{multline}
Now taking an expectation over a random choice of both
$\boldsymbol{\mathcal{V}}$ and $\boldsymbol{\mathcal{U}}$, by
(\ref{eq:uncond-typ}) and (\ref{eq:typicality}) and the gentle measurement
lemma \cite{W99,ON07}\ (see also \cite{W13}), it follows that%
\begin{equation}
\mathbb{E}_{\boldsymbol{\mathcal{V}},\boldsymbol{\mathcal{U}}}\left[
\operatorname{Tr}(\left[  I-\Gamma_{k,x}\right]  \mathcal{N}^{\otimes n}%
(\psi_{k}(x)))\right]  \leq2\sqrt{\varepsilon_{0}}+\varepsilon_{1}.
\end{equation}
Recalling that each unitary is independent of another, for $\left(  k^{\prime
},x^{\prime}\right)  \neq\left(  k,x\right)  $, we find that%
\begin{align}
& \mathbb{E}_{\boldsymbol{\mathcal{V}},\boldsymbol{\mathcal{U}}}\left[
\operatorname{Tr}(\Gamma_{k^{\prime},x^{\prime}}\mathcal{N}^{\otimes n}%
(\psi_{k}(x)))\right] \nonumber \\
&  =\operatorname{Tr}(\mathbb{E}_{\boldsymbol{\mathcal{V}}%
,\boldsymbol{\mathcal{U}}}\left[  \Gamma_{k^{\prime},x^{\prime}}\right]
\mathbb{E}_{\boldsymbol{\mathcal{V}},\boldsymbol{\mathcal{U}}}\left[
\mathcal{N}^{\otimes n}(\psi_{k}(x))\right]  )\\
&  =\operatorname{Tr}(\mathbb{E}_{\boldsymbol{\mathcal{V}}%
,\boldsymbol{\mathcal{U}}}\left[  \Gamma_{k^{\prime},x^{\prime}}\right]
\left[  \mathcal{N}(\pi)\right]  ^{\otimes n})\\
&  =\mathbb{E}_{\boldsymbol{\mathcal{V}},\boldsymbol{\mathcal{U}}%
}\operatorname{Tr}(\Pi_{k^{\prime},x^{\prime}}^{n,\delta_{1}}\Pi
_{\mathcal{N}(  \pi)  }^{n,\delta_{0}}\left[  \mathcal{N}%
(\pi)\right]  ^{\otimes n}\Pi_{\mathcal{N}(  \pi)  }^{n,\delta_{0}%
})\\
&  \leq2^{-\left[  S\left(  \mathcal{N}(  \pi)  \right)
-\delta_{0}\right]  }\operatorname{Tr}(\Pi_{k^{\prime},x^{\prime}}%
^{n,\delta_{1}}\Pi_{\mathcal{N}(  \pi)  }^{n,\delta_{0}})\\
&  \leq2^{-\left[  S\left(  \mathcal{N}(  \pi)  \right)
-\delta_{0}\right]  }\operatorname{Tr}(\Pi_{k^{\prime},x^{\prime}}%
^{n,\delta_{1}})\\
&  \leq2^{-\left[  S\left(  \mathcal{N}(  \pi)  \right)
-\delta_{0}\right]  }2^{n\left[  S_{\ast}+\delta_{1}\right]  }.
\end{align}
Putting everything together, we find that%
\begin{multline}
\mathbb{E}_{\boldsymbol{\mathcal{V}},\boldsymbol{\mathcal{U}}}\left[  \bar
{p}_{\operatorname{err}}\right]  \leq2\left[  2\sqrt{\varepsilon_{0}%
}+\varepsilon_{1}\right]  
\\+4KM2^{-\left[  S\left(  \mathcal{N}(  \pi)  \right)  -\delta
_{0}\right]  }2^{n\left[  S_{\ast}+\delta_{1}\right]  }.
\end{multline}
By picking%
\begin{equation}
\frac{1}{n}\log{M}+\frac{1}{n}\log{K}=S\left(  \mathcal{N}(  \pi)
\right)  -S_{\ast}-2\delta_{0}-2\delta_{1},\label{cor-final}%
\end{equation}
we can ensure that%
\begin{equation}
\mathbb{E}_{\boldsymbol{\mathcal{V}},\boldsymbol{\mathcal{U}}}\left[  \bar
{p}_{\operatorname{err}}\right]  \leq2\left[  2\sqrt{\varepsilon_{0}%
}+\varepsilon_{1}\right]  +4\cdot2^{-n\left(  \delta_{0}+\delta_{1}\right)
}.\label{eq:correct-err-bnd}%
\end{equation}
Eventually, we will derandomize the random bit-hiding code in order to
conclude the existence of one with this error probability bound.

\subsubsection{Security}

To simplify the notation we set $d = d_{B} d_{C}$. To prove the security of
the protocol against an LOCC measurement we show that
\begin{equation}
\frac{1}{M}\sum_{x}\left\Vert \mathcal{M}_{\operatorname{LOCC}}\left(
\mathcal{N}_{A\rightarrow BC}^{\otimes n}(\rho(x))-I^{\otimes n}/d^{n}\right)
\right\Vert _{1}\leq\varepsilon_{2}\,,
\end{equation}
for an arbitrarily small $\varepsilon_{2}\in\left(  0,1\right)  $ and for all
LOCC measurements. In fact, what we show is an even stronger bound by proving
that the protocol is secure against all separable measurements $\mathcal{M}%
_{\operatorname{sep}}$.


For a given codeword $\psi(x)$ and a separable measurement $\mathcal{M}%
_{\operatorname{sep}}$ (with POVM elements $\{B^{y}\otimes C^{y}\}$), we
define the conditional random variable $Y|\boldsymbol{\mathcal{U}}$ 
from the following distribution:%
\begin{equation}\label{pdistr}
p_{Y|\mathbf{U}}(y) = \operatorname{Tr}[\left(  B^{y}\otimes
C^{y}\right)  \mathcal{N}^{\otimes n}(\mathcal{R}_{\mathbf{U}}(\psi(x)))]\,.
\end{equation}

We then consider the mutual information of the random variables $Y$ and
$\boldsymbol{\mathcal{U}}$, given that codeword $\psi(x)$ was transmitted:%
\begin{equation}
\label{muinfo}I(Y;\boldsymbol{\mathcal{U}})=H(Y)-H(Y|\boldsymbol{\mathcal{U}})
\, .
\end{equation}


We first consider the case of separable projective measurements
$\mathcal{M}_{\operatorname{pro-sep}}$, which have separable POVM
elements $B^{y}\otimes C^{y} = \phi_{B}^{y} \otimes\phi_{C}^{y}$, where $\phi_{B}^{y}$ and
$\phi_{C}^{y}$ are rank-one projectors acting on Bob and Charlie's systems,
respectively.
By applying concentration inequalities to the conditional probability
distribution $p_{Y|\mathbf{U}}(y)$, we show that (see Appendix \ref{ci} for details)
for $K\gg K(n,d_{B},d_{C},\delta_{2})$%
\begin{equation}\label{IneedU}
\sup_{x,\mathcal{M}_{\operatorname{pro-sep}}} I(Y;\boldsymbol{\mathcal{U}%
})=O(\delta_{2}\log{d^{n}})\,,
\end{equation}
where $\delta_{2}\in\left(  0,1\right)  $ and
\begin{equation}
K(n,d_{B},d_{C},\delta_{2}) = 8 d_{+}^{n} \gamma^{n} \delta_{2}^{-2}\log
{\frac{10d^{n}}{\delta_{2}}}\,,
\end{equation}
with $d_{+} = \max{\{ d_{B} , d_{C}\}}$.

The Pinsker inequality (see, e.g., \cite{Pinsker2003}) states that 
\begin{align}
\| p_{Y,\mathbf{U}} - p_{Y} p_{\mathbf{U}} \|_1 
&= \int d\mathbf{U}\sum_{y}\left\vert p_{Y|\mathbf{U}}(y)-p_{Y}(y)\right\vert \nonumber \\
& \leq \sqrt{2\ln{2} I(Y;\boldsymbol{\mathcal{U}})} \, .
\label{PPinsker}
\end{align} 
Using (\ref{IneedU}) and the Pinsker inequality we then obtain
that for all $M$ codewords $\psi(x)$ and $\mathcal{M}_{\operatorname{pro-sep}}$
\begin{multline}
\int d\mathbf{U}\left\Vert \mathcal{M}_{\operatorname{pro-sep}} \left(
\mathcal{N}(\mathcal{R}_{\mathbf{U}}(\psi(x)))-I^{\otimes n}/d^{n}\right)
\right\Vert _{1} \\
= \int d\mathbf{U}\sum_{y}\left\vert p_{Y|\mathbf{U}}(y)-p_{Y}(y)\right\vert 
\leq  O(\sqrt{\delta_{2}\log{d^{n}}})\,.
\label{nnorms}
\end{multline}
Taking an average over all messages and exchanging this average with the
expectation over unitaries, we find that%
\begin{multline}
\mathbb{E}_{\boldsymbol{\mathcal{V}},\boldsymbol{\mathcal{U}}}\left\{
\frac{1}{M}\sum_{x}\left\Vert \mathcal{M}_{\operatorname{pro-sep}} \left(
\mathcal{N}(\mathcal{R}_{\mathbf{U}}(\psi(x)))-I^{\otimes n}/d^{n}\right)
\right\Vert _{1}\right\} \label{eq:hiding-bnd-last} \\
\leq O(\sqrt{\delta_{2}\log{d^{n}}}).
\end{multline}

Then, in Appendix \ref{meas} we show how this result can be extended to 
the case of generic separable measurements by applying some techniques 
developed in \cite{DFHL10}.

Finally, by choosing%
\begin{equation}
\frac{1}{n}\log{K}=\log{d_{+}}+\log{\gamma} +\lambda\frac{{\log n}}{n}\,,
\label{sec-final}%
\end{equation}
where $\lambda>0$ is a positive constant, the condition $K>K(n,d_{B}%
,d_{C},\delta_{2})$ is satisfied and implies that the LHS\ of
(\ref{eq:hiding-bnd-last}) decreases sub-exponentially in $n$.

\vspace{0.5cm}

To conclude the proof we combine the conditions (\ref{cor-final}) and
(\ref{sec-final}) to obtain that as long as the rate satisfies the following
condition%
\begin{multline}
\frac{1}{n}\log{M}=S\left(  \mathcal{N}(  \pi)  \right)  -S_{\ast
}-\log{d_{+}}\\-\log{\gamma}
-\lambda\frac{{\log n}}{n}-2\delta_{0}-2\delta_{1},
\end{multline}
then (\ref{eq:correct-err-bnd}) and (\ref{eq:hiding-bnd-last}) are satisfied.
This in turn implies that there exist choices of the unitaries $(U_{1}%
,\dots,U_{K})$ and $(V_{1},\dots,V_{M})$ such that these conditions are
satisfied and thus there exists an $\left(  n,M,\delta,\varepsilon\right)  $
bit-hiding code such that $\delta$ and $\varepsilon$ are vanishing with
increasing $n$. The asymptotic rate of the generated sequence of codes is then
as given in the statement of the theorem.

\subsection{Multipartite Generalization}

The result of Theorem \ref{main} is readily generalized to a multipartite
setting. Suppose for instance that Alice sends information from a system
of dimension $d_A$ through a mictodiactic channel $\mathcal{N}$ to $\ell$ 
receivers, $B_{1}, \dots, B_{\ell}$, with
Hilbert space dimensions $d_{j} = \dim{B_{j}}$, and we require security
against LOCC measurements. In this case we would obtain the following
achievable bit-hiding rate:
\begin{equation}
\chi(\mathcal{N}) - \log{d_{+}} - \log{\gamma} \,,
\end{equation}
with $d_{+} = \max_{j}{d_{j}}$ and 
\begin{equation}
\gamma = \frac{2 \prod_{j} d^2_{j} }{d_A( d_A+1)} \, \left\|
 \mathcal{N}^{\otimes 2}(P_\mathrm{sym}) \right\|  _{\infty} \, .
\end{equation}

\section{Discussion}

The main contribution of this paper is to provide a formal definition of the data hiding capacity of a quantum channel and upper and lower bounds on this operational quantity.
Our work thus initiates the study of ``noisy data hiding,'' and we now suggest several directions for future study. 

It could be possible to generalize Theorem~\ref{main} such that it applies to generic, non-mictodiactic, channels. 
Now we show how this generalization is obtained assuming two unproven statements.
We notice that the mictodiactic condition is used only to obtain (\ref{needed}), that is,
\begin{equation}
\mathbb{E}_{\boldsymbol{\mathcal{U}}}(X_{k})=\frac{1}{d^{n}}\,.
\end{equation}
where
\begin{equation}
X_{k}=\operatorname{Tr}\left(  \phi^{y}\mathcal{N}^{\otimes n}(U_{k}%
\psi(x)U_{k}^{\dag})\right)  \,.
\end{equation}
The vector $\phi^{y}$ corresponds to one of the POVM elements of a decoding
measurement.

The first unproven statement is that,
for $n$ large, $\phi^{y}$ belongs to the typical subspace associated to 
$\mathcal{N}^{\otimes n}(\mathbb{E}_{\boldsymbol{\mathcal{U}}}(U_{k}\psi(x)U_{k}^{\dag}))=\mathcal{N}^{\otimes n}(\pi_{A}^{\otimes n})=\mathcal{N}(\pi_{A})^{\otimes n}$, 
where $\pi_{A}=I_{A}/d_{A}$ denotes the maximally mixed state on system
$A$. Let  $\Pi_{\mathcal{N}(\pi_{A})}^{n,\delta}$ then denote the
projector onto the typical subspace. Putting $\phi^{y}=\Pi_{\mathcal{N}%
(\pi_{A})}^{n,\delta}\phi^{y}\Pi_{\mathcal{N}(\pi_{A})}^{n,\delta}$, we have
\begin{equation}
\mathbb{E}_{\boldsymbol{\mathcal{U}}}(X_{k})=\operatorname{Tr}\left(  \phi
^{y}\Pi_{\mathcal{N}(\pi_{A})}^{n,\delta}\mathcal{N}^{\otimes n}(\pi_{A}%
)\Pi_{\mathcal{N}(\pi_{A})}^{n,\delta}\right)  \,.
\end{equation}
The equipartition property of the typical subspace (see e.g.\ \cite{W13})
yields
\begin{multline}
2^{-n(S+\delta)}\Pi_{\mathcal{N}(\pi_{A})}^{n,\delta}\leq
\Pi_{\mathcal{N}(\pi_{A})}^{n,\delta}\mathcal{N}^{\otimes n}(\pi_{A}%
)\Pi_{\mathcal{N}(\pi_{A})}^{n,\delta} \\
\leq 2^{-n(S-\delta)}\Pi_{\mathcal{N}%
(\pi_{A})}^{n,\delta}\,,
\end{multline}
with $S=S(\mathcal{N}(\pi_{A}))$. This in turn implies
\begin{multline}
2^{-n(S+\delta)}=2^{-n(S+\delta)}\operatorname{Tr}\left(  \phi^{y}%
\Pi_{\mathcal{N}(\pi_{A})}^{n,\delta}\right)  \leq
\mathbb{E}_{\boldsymbol{\mathcal{U}}}(X_{k}) \\
\leq 2^{-n(S-\delta)}%
\operatorname{Tr}\left(  \phi^{y}\Pi_{\mathcal{N}(\pi_{A})}^{n,\delta}\right)
=2^{-n(S-\delta)} \,.
\end{multline}
These bounds generalize the condition (\ref{needed}), where $2^{S}$ replaces
$d_{B}d_{C}$ and plays the role of the effective dimension of the output system.

Similarly, consider the second moment, see (\ref{2mom}),
\begin{equation}
\mathbb{E}_{\boldsymbol{\mathcal{U}}}(X_{k}^{2}) = \mathrm{Tr} \left[
{\phi^{y}}^{\otimes2} \left(  \mathcal{N}^{\otimes2}\left(  \frac{P_\mathrm{sym}}{\mathrm{Tr(P_\mathrm{sym})}} \right)
\right)  ^{\otimes n} \right] \, .
\end{equation}

The second unproven statement is that
the measurement vectors are of the form
${\phi^{y}}^{\otimes2} = \Pi^{n,\delta}_{\mathcal{N}^{\otimes2}( P_\mathrm{sym} )}
{\phi^{y}}^{\otimes2} \Pi^{n,\delta}_{\mathcal{N}^{\otimes2}( P_\mathrm{sym} )}$, where
$\Pi^{n,\delta}_{\mathcal{N}^{\otimes2}( P_\mathrm{sym} )}$ is the typical projector
associated to the state $  \mathcal{N}^{\otimes2}\!\left( \frac{P_\mathrm{sym}}{\mathrm{Tr}P_\mathrm{sym}} \right) 
$. The equipartition property implies
\begin{multline}
 \mathrm{Tr} \left[  {\phi^{y}}^{\otimes2} \Pi^{n,\delta}_{\mathcal{N}^{\otimes
2}( P_\mathrm{sym} )} \left(  \mathcal{N}^{\otimes2}\left(  \frac{P_\mathrm{sym}}{\mathrm{Tr}P_\mathrm{sym}} \right)  \right)
^{\otimes n} \Pi^{n,\delta}_{\mathcal{N}^{\otimes2}( P_\mathrm{sym} )} \right] \\
 \leq 2^{-n(S_{2}-\delta)} \, ,
\end{multline}
where $S_{2} = S(\mathcal{N}^{\otimes2}( \frac{P_\mathrm{sym}}{\mathrm{Tr}P_\mathrm{sym}} ))$.

In conclusion, by modifying (\ref{gammabound}), it could be possible to 
extend Theorem~\ref{main} to non-mictodiactic (and non-unital) channels with $\gamma$ redefined as
\begin{equation}
\left(  \frac{\mathbb{E}_{\boldsymbol{\mathcal{U}}}(X_{k}^{2})}{\mathbb{E}%
_{\boldsymbol{\mathcal{U}}}(X_{k})^{2}}\right)  ^{\frac{1}{n}}\leq\left(
\frac{2^{-n(S_{2}-\delta)}}{2^{-2n(S+\delta)}}\right)  ^{\frac{1}{n}%
}=2^{2S-S_{2}+3\delta}=:\gamma\,.\label{gammabound-other}%
\end{equation}
Notice that this extension of Theorem \ref{main} also improves the bound for
mictodiactic channels. We leave a full development of the above observations for future work.

There are several ways in which the lower bound of Theorem \ref{main} could be
further improved. Let us recall that a map $\mathcal{R}$ is said to be an
approximately randomizing map with respect to a given norm $\Vert\cdot
\Vert_{\ast}$, if for all states $\psi$,
\begin{equation}
\Vert\mathcal{R}(\psi)-\pi\Vert_{\ast}\leq\varepsilon\,,
\end{equation}
where $\pi$ is the maximally mixed state. There exists a close relation
between data-hiding protocols and approximately randomizing maps. For
instance, the data-hiding protocol of \cite{HLSW04} was obtained by modifying
a related approximately randomizing map with respect to the operator norm
$\Vert\cdot\Vert_{\infty}$. Several constructions of approximately randomizing
maps with respect to the operator norm and the trace norm were discussed in
\cite{AS04,KN06,DN06,A09}. It could indeed be possible that these maps can be
used to develop quantum data hiding protocols robust to noise and achieving
higher rates. If this is true, it might be possible to achieve a bit-hiding
rate of%
\begin{equation}
\chi(\mathcal{N})-\log{d_{+}}\,.
\end{equation}

\textbf{Acknowledgements.} We thank Jonathan Olson, Graeme Smith, and Andreas
Winter for discussions about quantum data hiding capacity. MMW\ acknowledges
support from startup funds from the Department of Physics and Astronomy at LSU
and the NSF\ under Award No.~CCF-1350397. All authors acknowledge support from
the DARPA Quiness Program through US Army Research Office award W31P4Q-12-1-0019.

\appendices

\section{}

\label{sec:trace-dist-err-prob}

Let $M$ be a random variable. Let $p_{MM^{\prime}}$ denote the distribution in
which $M^{\prime}$ is a copy of $M$, so that%
\begin{equation}
p_{MM^{\prime}}(m,m^{\prime})=p_{M}(m)\delta_{m,m^{\prime}}.
\end{equation}
Let $q_{MM^{\prime}}$ denote the distribution that results from sending random
variable $M^{\prime}$ through a channel $q_{M^{\prime}|M}(m^{\prime}|m)$, so
that%
\begin{equation}
q_{MM^{\prime}}(m,m^{\prime})=p_{M}(m)q_{M^{\prime}|M}(m^{\prime}|m).
\end{equation}
Then the probability that $M^{\prime}\neq M$ under $q_{MM^{\prime}}$ is equal
to the normalized trace distance between $p_{MM^{\prime}}$ and $q_{MM^{\prime
}}$:%
\begin{equation}
\Pr_{q}\{M^{\prime}\neq M\}=\frac{1}{2}\left\Vert p_{MM^{\prime}%
}-q_{MM^{\prime}}\right\Vert _{1}.\label{eq:trace-dist-err-prob}%
\end{equation}
Consider that%
\begin{align}
\Pr_{q}\left\{  M^{\prime\prime}\neq M\right\}    & =\sum_{m^{\prime}\neq
m}q_{MM^{\prime}}(m,m^{\prime})\label{eq:err-prob-1}\\
& =\sum_{m^{\prime}\neq m}p_{M}(m)q_{M^{\prime}|M}(m^{\prime}%
|m).\label{eq:err-prob-2}%
\end{align}
Consider also that%
\begin{align}
& \left\Vert p_{MM^{\prime}}-q_{MM^{\prime}}\right\Vert _{1} \nonumber \\
& =\sum
_{m,m^{\prime}}\left\vert p_{M}(m)\delta_{m,m^{\prime}}-p_{M}(m)q_{M^{\prime
}|M}(m^{\prime}|m)\right\vert \label{eq:tr-dist-1}\\
& =\sum_{m,m^{\prime}}p_{M}(m)\left\vert \delta_{m,m^{\prime}}-q_{M^{\prime
}|M}(m^{\prime}|m)\right\vert \\
& =\sum_{m\neq m^{\prime}}p_{M}(m)q_{M^{\prime}|M}(m^{\prime}|m) \nonumber \\
& \qquad+\sum_{m}p_{M}(m)\left[  1-q_{M^{\prime}|M}(m|m)\right]  \\
& =2\sum_{m\neq m^{\prime}}p_{M}(m)q_{M^{\prime}|M}(m^{\prime}%
|m).\label{eq:tr-dist-last}%
\end{align}
So the equality in \eqref{eq:trace-dist-err-prob} follows from
\eqref{eq:err-prob-1}--\eqref{eq:err-prob-2} and \eqref{eq:tr-dist-1}--\eqref{eq:tr-dist-last}.

\section{Concentration inequalities}

\label{ci}In a crucial passage in the proof of Theorem \ref{main} we have
applied the bound
\begin{equation}
I(Y;\boldsymbol{\mathcal{U}})=O(\delta\log{d^{n}})\,,\label{AI}%
\end{equation}
holding for any separable measurement. Here and in Appendix \ref{meas} we
prove that this bound holds true provided
\begin{equation}
K\gg K(n,d_{B},d_{C},\delta)=8d_{+}^{n}\gamma^{n}\delta^{-2}\log{\frac
{10d^{n}}{\delta}}\,,
\end{equation}
with $\gamma$ as in (\ref{gammadef}).

We do so by first showing that (\ref{AI}) holds for separable projective
measurements $\mathcal{M}_{\operatorname{pro-sep}}$, which have separable POVM
elements $\phi_{B}^{y}\otimes\phi_{C}^{y}$, where $\phi_{B}^{y}$ and
$\phi_{C}^{y}$ are rank-one projectors acting on Bob and Charlie's systems,
respectively. This is done in Proposition \ref{MI}, where we apply techniques
analogous to those applied in \cite{LL14,LL15}. In Appendix \ref{meas} we show
how this result can be extended to the case of generic separable measurements
by applying some techniques developed in \cite{DFHL10}.

We will make use of two concentration inequalities. The first one is Maurer's tail
bound~\cite{M03}:

\begin{theorem}
\label{Maurer} Let $\{X_{k}\}_{k=1,\dots,K}$ be $K$ i.i.d.\ non-negative
real-valued random variables, with $X_{k}\sim X$ and finite first and second
moments: $\mathbb{E}[X],\mathbb{E}[X^{2}]<\infty$. Then, for any $\tau>0$ we
have that
\begin{equation}\label{MaurerINEQ}
\mathrm{Pr}\left\{  \frac{1}{K}\sum_{k=1}^{K}X_{k}<\mathbb{E}[X]-\tau\right\}
\leq\exp{\left(  -\frac{K\tau^{2}}{2\mathbb{E}[X^{2}]}\right)  }\,.
\end{equation}

\end{theorem}

The second one is the Chernoff bound~\cite{AW02}:

\begin{theorem}
\label{Chernoff} Let $\{X_{k}\}_{k=1,\dots,K}$ be $K$ i.i.d.\ random
variables, $X_{k}\sim X$, with $0\leq X\leq1$ and $\mathbb{E}[X]=\mu$. Then,
for any $\tau>0$ such that $(1+\tau)\mu\leq1$ we have that
\begin{equation}
\mathrm{Pr}\left\{  \frac{1}{K}\sum_{k=1}^{K}X_{k}>(1+\tau)\mu\right\}
\leq\exp{\left(  -\frac{K\tau^{2}\mu}{4\ln{2}}\right)  }\,.
\end{equation}

\end{theorem}

For given states $\psi(x)$ and $\phi^{y}=\phi_{B}^{y}\otimes\phi_{C}^{y}$,
we apply these bounds to the quantity
\begin{equation}
X_{k}=\operatorname{Tr}\left(  \phi^{y}\mathcal{N}^{\otimes n}(U_{k}%
\psi(x)U_{k}^{\dag})\right)  \,.
\end{equation}
This quantity is a random variable if the unitary $U_{k}$ is chosen randomly.
Taking the average over the random unitaries $U_{k}$ we obtain
\begin{equation}
\mathbb{E}_{\boldsymbol{\mathcal{U}}}(U_{k}\psi(x)U_{k}^{\dag})=\frac
{I^{\otimes n}}{d_{A}^{n}}\,,
\end{equation}
which for mictodiactic channels implies
\begin{equation}
\mathbb{E}_{\boldsymbol{\mathcal{U}}}(X_{k})=\frac{ 1 }{ d_B^{n} d_C^{n} } \,. \label{needed}
\end{equation}

We also consider the second moment
\begin{multline}\label{IImoment}
\mathbb{E}_{\boldsymbol{\mathcal{U}}}(X_{k}^{2})=\int dU_{k}\operatorname{Tr}%
\left[  \left(  \phi^{y}\mathcal{N}^{\otimes n}(U_{k}\psi(x)U_{k}^{\dag
})\right)  ^{\otimes2}\right]  \\
=\mathrm{Tr}\left[  {\phi^{y}}^{\otimes2}\bigotimes_{j=1}^{n}\mathcal{N}^{\otimes
2}\left(  \int dU_{kj}\,U_{kj}^{\otimes2}\psi_j(x)^{\otimes2}{U_{kj}^{\dag}%
}^{\otimes2}\right)  \right]  \,.
\end{multline}
For any two-qudit pure state $\rho$ we have 
\begin{equation}
\int dU\,U^{\otimes 2}\rho^{\otimes2}{U^{\dag}}^{\otimes2}=\frac{2}{d_A(d_A+1)}P_\mathrm{sym} \,, 
\end{equation}
where $P_\mathrm{sym}$ is the projector onto the symmetric subspace (see, e.g., \cite{H13}). We then
obtain
\begin{align}
\mathbb{E}_{\boldsymbol{\mathcal{U}}}(X_{k}^{2}) & = \left[\frac{2}{d_A(d_A+1)}\right]^n \mathrm{Tr}\left[  
{\phi^{y}}^{\otimes2}\left(  \mathcal{N}^{\otimes2}\left(  P_\mathrm{sym}\right)  \right)
^{\otimes n}\right]  \\
& \leq \left[\frac{2}{d_A(d_A+1)}\right]^n \Vert\mathcal{N}^{\otimes2}\left(  P_\mathrm{sym}\right)
\Vert_{\infty}^{n} \label{2mom}%
\end{align}
(where we have used the fact that the expectation value of an operator on a vector is 
upper bounded by the $\infty$-norm), from which it follows
\begin{equation}
\left(  \frac{\mathbb{E}_{\boldsymbol{\mathcal{U}}}(X_{k}^{2})}{\mathbb{E}%
_{\boldsymbol{\mathcal{U}}}(X_{k})^{2}}\right)  ^{\frac{1}{n}}\leq \frac{2 d_B^2 d_C^2}{d_A(d_A+1)}%
\Vert\mathcal{N}^{\otimes2}\left(  P_\mathrm{sym} \right)  \Vert_{\infty}=\gamma
\,.\label{gammabound}%
\end{equation}

We will also make use of the notion of a net of unit vectors \cite{HLSW04}.
Let us recall that an $\varepsilon$-net is a finite set of unit vectors
$N_{\varepsilon}=\{\tilde{\phi}\}$ in a $D$-dimensional Hilbert space such
that for any unit vector $\phi$ there exists $\widetilde{\phi}\in
N_{\varepsilon}$ for which
\begin{equation}
\Vert\phi-\widetilde{\phi}\Vert_{1}\leq\varepsilon\,.
\end{equation}
As discussed in~\cite{HLSW04} there exist $\varepsilon$-nets with less than
$(5/\varepsilon)^{2D}$ elements.
%
Let us consider a pair of vectors ${\phi_{B}}$, ${\phi_{C}}$ and the
corresponding vectors $\widetilde{\phi}_{B}$, $\widetilde{\phi}_{C}$ in the
net such that
\begin{equation}
\Vert{\phi_{B}}-\widetilde{\phi}_{B}\Vert_{1}\,,\Vert{\phi_{C}}-\widetilde
{\phi}_{C}\Vert_{1}\leq\varepsilon\,.
\end{equation}
We consider the quantity 
defined in (\ref{pdistr}):%
\begin{equation}
p_{Y|\mathbf{U}}(y) 
= \operatorname{Tr}[\left(  \phi_{B}^y \otimes \phi_{C}^y \right)  \mathcal{N}^{\otimes n}(\mathcal{R}_{\mathbf{U}%
}(\psi(x)))]\,,
\end{equation}
and
\begin{equation}
\tilde p_{Y|\mathbf{U}}(y) 
= \operatorname{Tr}[\left(  {\tilde \phi}_{B}^y \otimes {\tilde \phi}_{C}^y \right)  \mathcal{N}^{\otimes n}(\mathcal{R}_{\mathbf{U}%
}(\psi(x)))]\,.
\end{equation}

We have
\begin{equation}
\left\vert p_{Y|\mathbf{U}}(y) - \tilde p_{Y|\mathbf{U}}(y) \right\vert \leq 2 \varepsilon\,.
\end{equation}
Finally, we consider the function $\eta(\,\cdot\,)=-(\,\cdot\,)\log
{(\,\cdot\,)}$. Using the inequality $\left\vert \eta(r)-\eta(s)\right\vert
\leq\eta(|r-s|)$ (which holds for $\left\vert r-s\right\vert <1/2$) we obtain
\begin{equation}
\left\vert \eta(p_{Y|\mathbf{U}}(y)) - \eta(\tilde p_{Y|\mathbf{U}}(y))\right\vert \leq\eta
(2\varepsilon)\,.\label{FA}%
\end{equation}

\vspace{0.5cm}

From now on, to make notation lighter, we put $d:= d_B d_C$.
We are now ready to prove the following:

\begin{proposition}
\label{MI} For any $\delta\in\left(  0,1\right)  $ and $K>K(n,d_{B}%
,d_{C},\delta)$,
\begin{equation}
\sup_{x,\mathcal{M}_{\operatorname{pro-sep}}}I(Y;\boldsymbol{\mathcal{U}}) =
O( \delta\log{d^{n}} ) \,,
\end{equation}
where the sup is over separable projective measurements $\mathcal{M}%
_{\operatorname{pro-sep}}$.
\end{proposition}

First of all, we notice that for a projective measurement with associated POVM
elements $\{\phi^{y}=\phi_{B}^{y}\otimes\phi_{C}^{y}\}_{y=1,\dots,d^{n}}$
the mutual information (\ref{muinfo}) reads
\begin{equation}
I(Y;\boldsymbol{\mathcal{U}}) = \log{d^{n}} - H(Y|\boldsymbol{\mathcal{U}}) \,
,
\end{equation}
where we have also used the condition (\ref{needed}).

For a given vector $\psi(x)$ and separable projective measurement
$\mathcal{M}_{\operatorname{pro-sep}}$ we consider the quantity
\begin{align}
H(Y)_{\mathbf{U}}  & =-\sum_{y}p_{Y|\mathbf{U}}(y)\log{p_{Y|\mathbf{U}}
(y)} \\
& =\sum_{y}\eta\lbrack p_{Y|\mathbf{U}}(y)] \, .
\end{align}

For a given $y$, we want to bound the probability that $H(Y)_{\mathbf{U}} \leq (1-\delta)\log{d^{n}}$.
In order to do that we bound the probability that
$\eta \lbrack p_{Y|\mathbf{U}}(y)] \leq \eta \lbrack \frac{1-\delta}{d^{n}} ]$.
Notice that $\eta \lbrack x ]$ is a concave function and the equation 
$\eta \lbrack x ] = \eta \lbrack \frac{1-\delta}{d^{n}}]$ has two roots:
$x_- = \frac{1-\delta}{d^{n}}$, and $x_+ = 1 - \eta \lbrack \frac{1-\delta}{d^{n}}] + O(\eta \lbrack \frac{1-\delta}{d^{n}}])$,
where for $d^n$ large enough
$x_+ \geq 1 - 2\eta \lbrack \frac{1-\delta}{d^{n}}]$.
Therefore we have
\begin{multline}
\mathrm{Pr}_{\boldsymbol{\mathcal{U}}}\left\{  \eta\lbrack p_{Y|\mathbf{U}%
}(y)]\leq\frac{1-\delta}{d^{n}}\log{d^{n}}\right\} \\
\leq\mathrm{Pr}_{\boldsymbol{\mathcal{U}}}\left\{  p_{Y|\mathbf{U}}%
(y)\leq x_- \right\} 
+\mathrm{Pr}_{\boldsymbol{\mathcal{U}}}\left\{  p_{Y|\mathbf{U}}%
(y)\geq x_+\right\} \\
\leq\mathrm{Pr}_{\boldsymbol{\mathcal{U}}}\left\{  p_{Y|\mathbf{U}}%
(y)\leq\frac{1-\delta}{d^{n}}\right\} 
\\+\mathrm{Pr}_{\boldsymbol{\mathcal{U}}}\left\{  p_{Y|\mathbf{U}}%
(y)\geq1-\delta^{\prime}\right\}  \,,
\end{multline}
where $\delta^{\prime}\equiv 2\eta \lbrack \frac{1-\delta}{d^{n}}]$.

For given $x$ and $y$, we now apply the Maurer tail bound (Theorem \ref{Maurer}) to the random variables
\begin{equation}
X_k := \operatorname{Tr}\left(
\phi_{B}^{y}\otimes\phi_{C}^{y} \, \mathcal{N}^{\otimes n}(U_{k}%
\psi(x)U_{k}^{\dag}) \right)  \, ,
\end{equation}
whose first and second moments are given by  (\ref{needed}) and (\ref{IImoment})
and obey the inequality in (\ref{gammabound}).
We remark that 
\begin{equation}
\frac{1}{K} \sum_{k} X_k  = p_{Y|\boldsymbol{\mathbf{U}}}(y) \, .
\end{equation}
Hence, applying (\ref{MaurerINEQ}) with $\tau = \delta \mathbb{E}[X] = \delta/d^n$,
we obtain
\begin{align}
\mathrm{Pr}_{\boldsymbol{\mathcal{U}}} \left\{  p_{Y|\boldsymbol{\mathbf{U}}}(y) < \frac{1-\delta}{d^n} \right\}
& \leq \exp{\left(  -\frac{K \delta^2 \mathbb{E}[X]^{2}}{2\mathbb{E}[X^{2}]}\right)  } \\
& \leq \exp{\left(  -\frac{K\delta^{2}}{2\gamma^{n}}\right)  } \,.
\end{align}

Similarly we apply the Chernoff inequality, Theorem \ref{Chernoff}, and
obtain
\begin{multline}
\mathrm{Pr}_{\boldsymbol{\mathcal{U}}}\left\{  p_{Y|\mathbf{U}}(y)\geq
1-\delta^{\prime}\right\}  
\leq 
\exp{\left(  -\frac{Kd^{n}(1-\delta'-1/d^n)^{2}}{4\ln{2}}\right)  } \\
\leq 
\exp{\left(  -\frac{Kd^{n}(1/2)^2}{4\ln{2}}\right)  } \, ,
\end{multline}
where the last inequality holds for sufficiently small $\delta'$ and large $d^n$.
We then have
\begin{align}
& \mathrm{Pr}_{\boldsymbol{\mathcal{U}}}\left\{  \eta(p_{Y|\mathbf{U}%
}(y))\leq\frac{1-\delta}{d^{n}}\log{d^{n}}\right\} \nonumber \\
&  \leq\exp{\left(  -\frac{K\delta^{2}}{2\gamma^{n}}\right)  } + \exp{\left(
-\frac{Kd^{n}(1/2)^{2}}{4\ln{2}}\right)  }\\
&  \leq2 \exp{\left(  -\frac{K\delta^{2}}{2\gamma^{n}}\right)  } \,,
\end{align}
where the last inequality holds for any $d^n > 8 \ln{2} \gamma^{-n} \delta^2$.

This is true for given $x$ and vectors $\phi_{B}^{y}$, $\phi_{C}^{y}$. To
account for all possible codewords $\psi(x)$ and measurement vectors
$\phi_{B}^{y}$, $\phi_{C}^{y}$, we introduce a $(\delta2^{-1}d^{-n})$-net
for Bob's system and a $(\delta2^{-1}d^{-n})$-net for Charlie's one,
containing in total no more than $\left(  10d^{n}/\delta\right)  ^{2d_{B}%
^{n}+2d_{C}^{n}}$ elements. Applying the union bound on the net's vectors and
on the $M$ codewords we obtain
\begin{align}
& \mathrm{Pr}\left\{  \inf_{x,\tilde{\phi}_{B}^{y},\tilde{\phi}_{C}^{y}} \eta(p_{Y|\mathbf{U}}(y))\leq\frac{1-\delta}{d^{n}}\log{d^{n}}\right\} \nonumber \\
&  \leq2 M \left(  \frac{10d^{n}}{\delta}\right)  ^{2d_{B}^{n}+2d_{C}^{n}}
\exp{\left(  -\frac{K\delta^{2}}{2\gamma^{n}}\right)  } \\
&  \leq2M\left(  \frac{10d^{n}}{\delta}\right)  ^{4d_{+}^{n}}\exp{\left(
-\frac{K\delta^{2}}{2\gamma^{n}}\right)  } \\
&  \leq\exp{\left(  -\frac{K\delta^{2}}{2\gamma^{n}}+4d_{+}^{n}\log
{\frac{10d^{n}}{\delta}}+\log{(2M)}\right)  } \\
&  =:p\,,
\end{align}
where $d_{+}=\max\{d_{B},d_{C}\}$. From this, we can extend the infimum over
all unit vectors by paying a small penalty given by (\ref{FA}). In this way we
obtain
\begin{equation}
  \mathrm{Pr}\Bigg\{  \inf_{\phi_{B}^{y},\phi_{C}^{y}}\eta
(p_{Y|\mathbf{U}}(y)) 
\leq\frac{1-\delta}{d^{n}}\log{d^{n}} 
    - \eta(\delta d^{-n}) \Bigg\}  \leq p \, .
\end{equation}
The probability of such a bad event can be made arbitrarily small provided (we
are assuming $\log{(2M)}\ll4d_{+}^{n}\log{\frac{10d^{n}}{\delta}}$)
\begin{equation}
K \gg8 \gamma^{n}d_{+}^{n}\delta^{-2}\log{\frac{10d^{n}}{\delta}}\,.
\end{equation}
Under this condition we then have that, up to a small probability $p$, for all
separable projective measurements $\mathcal{M}_{\operatorname{pro-sep}}$,
\begin{equation}
H(Y)_{\mathbf{U}} \geq(1-\delta)\log{d^{n}} - d^{n} \eta(\delta d^{-n}) \,.
\end{equation}
This implies
\begin{align}
 H(Y|\boldsymbol{\mathcal{U}}) & = \int d\mathbf{U} \, H(Y)_{\mathbf{U}} \\
&  \geq\int d\mathbf{U} \, H(Y)_{\mathbf{U}} \, \Theta \\
&  \geq\left[  (1-\delta)\log{d^{n}} - d^{n} \eta(\delta d^{-n}) \right]
(1-p) \, ,
\end{align}
where $\Theta\equiv\Theta\left[  H(Y)_{\mathbf{U}} - (1-\delta)\log{d^{n}} +
d^{n} \eta(\delta d^{-n}) \right]  $ denotes the Heaviside step function,
which finally yields
\begin{multline}
\sup_{x,\mathcal{M}_{\operatorname{pro-sep}}}I(Y;\boldsymbol{\mathcal{U}})
\leq\delta\log{d^{n}} + d^{n} \eta(\delta d^{-n})\\
+ \left[  (1-\delta)\log{d^{n}} - d^{n} \eta(\delta d^{-n}) \right]  p\\
= O\left(  \delta\log{d^{n}} \right)  \, .
\end{multline}

\section{From projective measurements to generic separable measurements}\label{meas}

In this appendix we discuss how Proposition \ref{MI} can be
extended to include generic separable measurements via the notion of 
{\it separable quasi-measurements}. 
In order to do that we apply a simple modification of Lemma 6.2 in \cite{DFHL10}.

The notion of quasi-measurement was defined in \cite{DFHL10} :
\begin{definition}[Quasi-Measurement \cite{DFHL10}]
\label{def:quasiM}
We call $(s,f)$ quasi-measurement an incomplete measurement on a
$D$-dimensional system such that the associated POVM elements are
of the form $\frac{D^2}{s} \, \phi^{y}$, where
$ \phi^{y}$, for $y=1,\dots,s$, are rank-one projectors and
$\sum_{y=1}^{s}\frac{D^2}{s} \phi^{y} \leq f \, I$. 
We denote as $\mathcal{L}(s,f)$ the set of $(s,f)$-separable quasi-measurements.
\end{definition}

Lemma 6.2 in \cite{DFHL10} proves that quasi-measurements are 
almost as informative as generic measurements, that is,
given a bipartite quantum state $\rho_{UB}$ we have 
\begin{multline}
\sup_\mathcal{M} \left\| \mathcal{M}\left( \rho_{UB} - \rho_U \otimes \rho_B \right) \right\|_1
\\\leq \sup_{\mathcal{M}' \in \mathcal{L}(s,f)} \left\| \mathcal{M}'\left( \rho_{UB} - \rho_U \otimes \rho_B \right) \right\|_1 
\\+ 4 D^2 e^{-s(\eta-1)^2/(D(2\ln{2}))} \, ,
\end{multline}
where the $\sup$ on left hand side is over generic measurements
$\mathcal{M}_{B \to Y}$ on the $B$ system, 
and that on the right hand side is over quasi-measurements
$\mathcal{M}'_{B \to Y}$.


To apply this result to our setting, we first define the notion of separable
quasi-measurements by adapting Definition \ref{def:quasiM} to the LOCC setting:
\begin{definition}\label{def:sep-quasiM}
[Separable quasi-measurement]
We say that an incomplete measurement on a
$D_{B} \times D_{C}$-dimensional bipartite system is an $(s,f)$-separable quasi-measurement 
if the associated POVM elements have the form
$\frac{ ( D_{B} D_{C} )^2 }{s}\phi_{B}^{y}\otimes\phi_{C}^{y}$, where
$\phi_{B}^{y}$, $\phi_{C}^{y}$, for $y=1,\dots,s$, are rank-one
projectors, such that $\sum_{y=1}^{s}\frac{ (D_{B} D_{C})^{2}}{s}\phi_{B}^{y}\otimes \phi_{C}^{y}\leq f \, I_{BC}$. 
We denote as $\mathcal{L}_{\mathrm{sep}}(s,f)$ the set of $(s,f)$-separable quasi-measurements.
\end{definition}

We notice that the proof in Appendix \ref{ci}, which considers separable
projective measurements, also applies to the case of separable
quasi-measurements. This implies that, by considering the set of $(s,f)$-separable
quasi-measurements, a modified version of Proposition \ref{MI} will be obtained, i.e.,
\begin{equation}
\sup_{x,\mathcal{M}\in\mathcal{L}_{\mathrm{sep}}(s,f)}%
I(Y;\boldsymbol{\mathcal{U}})=O(\delta\log{s}).
\end{equation}

To move from separable quasi-measurements to generic separable measurements we apply a
straightforward modification of Lemma 6.2 in \cite{DFHL10} in the LOCC setting, that is,
\begin{multline}
\sup_{\mathcal{M}_\mathrm{sep}} \left\| \mathcal{M}\left( \rho_{UBC} - \rho_U \otimes \rho_{BC} \right) \right\|_1
\\\leq \sup_{\mathcal{M}'^\mathrm{sep} \in \mathcal{L}_\mathrm{sep}(s,f)} \left\| \mathcal{M}'\left( \rho_{UBC} - \rho_U \otimes \rho_{BC} \right) \right\|_1 
\\+ 4 (D_B D_C)^2 e^{-s(\eta-1)^2/(D_B D_C(2\ln{2}))} \, ,
\label{finalone}
\end{multline}
where the $\sup$ on the left hand side is over separable measurements
$\mathcal{M}^\mathrm{sep}_{B,C \to Y}$, and the one on the right is over 
separable quasi-measurements $\mathcal{M}'^{\mathrm{sep}}_{B,C \to Y}$.


In conclusion, to extend Proposition \ref{MI} to generic separable measurements 
we proceed as follows.
First recall that (see Eqs.\ (\ref{PPinsker}) and (\ref{nnorms})) 
given a separable measurement $\mathcal{M}_\mathrm{sep}$, in our setting we have
\begin{multline}
\left\| \mathcal{M}_\mathrm{sep} \left( \rho_{UBC} - \rho_U \otimes \rho_{BC} \right) \right\|_1 = \\ 
\int d\mathbf{U}\left\Vert \mathcal{M}_{\operatorname{sep}} \left(
\mathcal{N}(\mathcal{R}_{\mathbf{U}}(\psi(x)))-I^{\otimes n}/d^{n}\right)
\right\Vert _{1} \,.
\end{multline}
Then consider that Proposition \ref{MI}, together with (\ref{PPinsker}), states that
\begin{multline}
\sup_{\mathcal{M}_\mathrm{pro-sep}} \left\| \mathcal{M}_\mathrm{pro-sep} \left( \rho_{UBC} - \rho_U \otimes \rho_{BC} \right) \right\|_1 \\
\leq O(\sqrt{\delta \log{d^{n}}}) \, ,
\end{multline}
where the $\sup$ is over projective separable measurements. 
As noticed above, Proposition \ref{MI} is straightforwardly extended from
projective separable measurements to separable quasi-measurements, yielding
\begin{multline}
\sup_{ \mathcal{M}\in\mathcal{L}_{\mathrm{sep}}(s,f) } 
\left\| 
\mathcal{M} \left( \rho_{UBC} - \rho_U \otimes \rho_{BC} \right) 
\right\|_1 
\leq O(\sqrt{\delta \log{s}}) \, .
\end{multline}
Finally, using (\ref{finalone}) with $D_B = d_B^n$, $D_C = d_C^n$, we obtain 
\begin{multline}
\sup_{\mathcal{M}_\mathrm{sep}} \left\| \mathcal{M}\left( \rho_{UBC} - \rho_U \otimes \rho_{BC} \right) \right\|_1
\\ \leq O(\sqrt{\delta \log{s}}) + 4 (d_B d_C)^{2n} e^{-s(\eta-1)^2/(d_B^n d_C^n(2\ln{2}))} \, .
\end{multline}

Putting, for example, $s=d^{2n}$ and $f=1/2$, yields
\begin{multline}
\sup_{\mathcal{M}_\mathrm{sep}} \left\| \mathcal{M}\left( \rho_{UBC} - \rho_U \otimes \rho_{BC} \right) \right\|_1
\\ \leq O(\sqrt{\delta \log{d^{2n}}})
 + 4d^{2n}\exp{\left(  -\frac{d^{n}}{8\ln{2}}\right)  } \,.
\end{multline}
This result can then be used to extend the security to the case of 
generic separable measurements by paying a small penalty not larger than
$4d^{2n}\exp{\left(  -\frac{d^{n}}{8\ln{2}}\right)  }$.

\bibliography{ref}{}
\bibliographystyle{IEEEtran}

\vspace{0.5cm}

\begin{IEEEbiographynophoto}{Cosmo Lupo}
received the Ph.D. degree in Fundamental and Applied Physics 
from the University of Napoli ``Federico II''. He has been a postdoctoral researcher
at the University of Camerino and at the Massachusetts Institute of Technology.
His research interests include quantum communication theory, quantum metrology, 
and quantum computing.
\end{IEEEbiographynophoto}

\begin{IEEEbiographynophoto}{Mark M.~Wilde} (M'99--SM'13) was born in Metairie, Louisiana, USA. 
He is an Assistant Professor in the Department of Physics and Astronomy and the Center for Computation and 
Technology at Louisiana State University. His current research interests are in quantum Shannon theory, 
quantum optical communication, quantum computational complexity theory, and quantum error correction.
\end{IEEEbiographynophoto}

\begin{IEEEbiographynophoto}{Seth Lloyd} is Nam P. Suh Professor of Mechanical Engineering and Professor of Physics at the Massachusetts
Institute of Technology, and on the external faculty of the Santa Fe Institute. His work focuses on quantum information
theory, including quantum communications, quantum algorithms, quantum metrology, and methods for building
quantum computers.
\end{IEEEbiographynophoto}

\end{document}